# Entanglement: Balancing Punishment and Compensation, Repeated Dilemma Game-Theoretic Analysis of Maximum Compensation Problem for Bypass and Least Cost Paths in Fact-Checking, Case of Fake News with Weak Wallace's Law

Yasuko Kawahata [†]

Faculty of Sociology, Department of Media Sociology, Rikkyo University, 3-34-1 Nishi-Ikebukuro,Toshima-ku, Tokyo, 171-8501, JAPAN.

ykawahata@rikkyo.ac.jp

**Abstract:** This research note is organized with respect to a novel approach to solving problems related to the spread of fake news and effective fact-checking. Focusing on the least-cost routing problem, the discussion is organized with respect to the use of Metzler functions and Metzler matrices to model the dynamics of information propagation among news providers. With this approach, we designed a strategy to minimize the spread of fake news, which is detrimental to informational health, while at the same time maximizing the spread of credible information. In particular, through the punitive dominance problem and the maximum compensation problem, we developed and examined a path to reassess the incentives of news providers to act and to analyze their impact on the equilibrium of the information market. By applying the concept of entanglement to the context of information propagation, we shed light on the complexity of interactions among news providers and contribute to the formulation of more effective information management strategies. This study provides new theoretical and practical insights into issues related to fake news and fact-checking, and will be examined against improving informational health and public digital health.

## 1. Introduction

This research note organizes the discussion regarding the development of an agent model that combines the iterated prisoner's dilemma game, the peer effect, and the neighbor matrix as a new approach to eliminate the filter bubble. The model aims to design strategies to mitigate information bias and echo chamber phenomena and promote a healthier information environment. In addition, by incorporating the concepts of crude substitutability (weak Walras' law) and least-cost paths of circumvention, we will consider with respect to a game-theoretic analysis of the complex dynamics of fake news proliferation and its punishment and compensation.

### 1.1 Incorporating Crude Substitutability (Weak Walras Law)

By incorporating crude substitutability, or weak Walras' law, into the model, we apply the economic principle that as the "price" of fake news increases (i.e., as the spread of fake news becomes more costly due to punishment), agents increase their reliance on truthful information and reliable sources

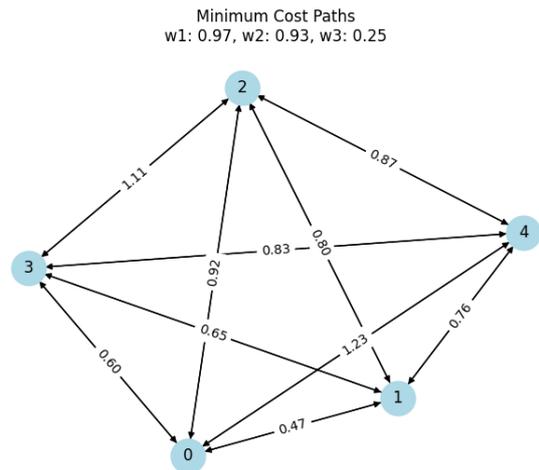

Fig. 1: Minimum Cost Path, Information Value and Cost / Demand-Supply Equilibrium

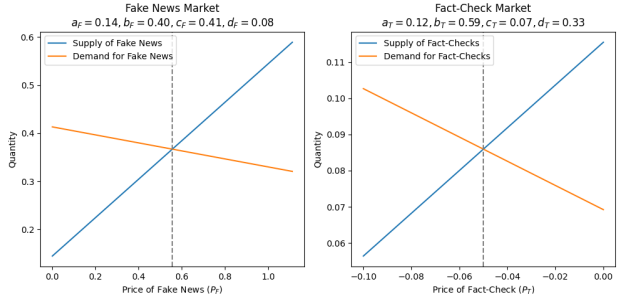

Fig. 2: Price of Fact-Check, Price of Fake News

This is a good thing. In doing so, we analyze the strategic interactions among information providers and the supply-demand balance in the information market, and consider the "cost" perspective on promoting a healthy information environment.

### 1.2 Least cost paths of detour routes, Metzler functions and Metzler matrices

We aim to optimize information propagation by introducing the concept of least cost paths of detour routes. We will use Metzler functions and Metzler matrices to mathematically model the paths of information propagation and their costs, and consider strategies to identify and minimize the spread of fake news within the filter bubble, and consider solution methods that introduce Metzler functions and Metzler matrices based on agent models.

### 1.3 Punitive Dominance Problem and Maximum Compensation Problem

We analyze the Punitive Dominance Problem, in which the punishment for the spreader of fake news exceeds the gain from the act of spreading it, and the Maximum Compensation Problem, in which sufficient incentives are provided to the providers of reliable information and agents who contribute to fact-checking. Game-theoretic analysis of these problems will also be examined to design strategies to prevent the spread of fake news and promote the sharing of truthful information.

The model uses an iterated prisoner's dilemma game to simulate the process of agents updating their strategies based on their history of past interactions. By considering multiple strategies, including tight-trigger and grim-trigger strategies, and by analyzing peer effects and neighbor matrices among agents, we obtain gain functions, expected gain functions, and also crude substitutability (weak Walras' law) with respect to agents' behavioral tendencies to spread fake news and contribute to fact checking The incorporation also touches on the "cost" issue of the concept of least cost paths of detour routes.

This analytical approach will allow us to design strategies to eliminate the filter bubble and promote a healthier and more diverse information environment. This will be an introduction to organizing the concept of designing a balance between appropriate punishment for the spreaders of fake news and appropriate compensation for agents who provide truthful information.

## 2. Discussion:Substitutability (Weak Walras' Law)

The concept of gross substitutability (weak Walras' law) is one of the theories in economics, particularly concerning the balance of supply and demand in markets. This law states that when the total supply of all goods in the market equals the total demand for those goods, the entire market is in equilibrium. When applied in the context of information markets, different types of information (e.g., fake news and fact-checking information) are treated as goods, and the analysis examines how each supply and demand contributes to the equilibrium of the information market.

**Theoretical Explanation and Computational Process**
**Market Definition**
1. Consider an information market with $N$ different types of information goods denoted as $i$, each with a price $p_i$.

**Supply and Demand Functions**
2. Define supply function $S_i(p_i)$ and demand function $D_i(p_i)$ for each information good $i$. These are expressed as functions of price $p_i$.

$$S_i(p_i) = a_i + b_i p_i$$

$$D_i(p_i) = c_i - d_i p_i$$

Here, $a_i, b_i, c_i, d_i$ are parameters, with $b_i, d_i > 0$.
**Equilibrium Condition**
3. The equilibrium condition in the market is that supply equals demand for each information good.

$$S_i(p_i) = D_i(p_i) \quad \forall i$$

**Application of Weak Walras' Law**
4. According to weak Walras' law, when the equilibrium condition holds for all information goods, the entire market is considered to be in equilibrium. If the total supply equals the total demand in the market, then all markets are in equilibrium.

$$\sum_{i=1}^{N} p_i S_i(p_i) = \sum_{i=1}^{N} p_i D_i(p_i)$$

**Calculation of Equilibrium Prices**
5. To find the prices $p_i$ that satisfy the equilibrium condition, solve the equations of supply and demand functions. This is generally represented in the form of simultaneous equations and often solved numerically.

$$a_i + b_i p_i = c_i - d_i p_i \quad \Rightarrow \quad p_i = \frac{c_i - a_i}{b_i + d_i}$$

**Analysis of Equilibrium State**

6. Using the obtained equilibrium prices $p_i$, calculate the quantities supplied and demanded for each information good and analyze the overall equilibrium state of the market.

$$S_i^* = S_i(p_i^*) \quad \text{and} \quad D_i^* = D_i(p_i^*)$$

Here, $p_i^*$ represents the equilibrium price, and $S_i^*$ and $D_i^*$ represent the quantities supplied and demanded at equilibrium.

Through this theoretical explanation and computational process, one can understand the circulation and equilibrium state of fake news and fact-checking information in information markets and contemplate strategies to improve informational health.

When considering the issue of the spread of fake news and the cost of fact-checking within the framework of gross substitutability (weak Walras' law), it is important to analyze the equilibrium of supply and demand in the information market. In this approach, the costs of spreading fake news and fact-checking function as "prices" in the information market and are believed to influence the behavior choices of news providers.

**Proposal of Formulas and Computational Process**

**Definition of Information Market**

Define the supply functions for fake news $F$ and fact-checking information $T$.

$$S_F(P_F) = a_F + b_F \cdot P_F$$

$$S_T(P_T) = a_T + b_T \cdot P_T$$

Here, $P_F$ and $P_T$ are the market prices for fake news and fact-checking information respectively, and $a_F, b_F, a_T, b_T$ are parameters. Similarly, define the demand functions for fake news $F$ and fact-checking information $T$.

$$D_F(P_F) = c_F - d_F \cdot P_F$$

$$D_T(P_T) = c_T - d_T \cdot P_T$$

Here, $c_F, d_F, c_T, d_T$ are parameters.

**Calculation of Equilibrium Prices**

Calculate the equilibrium prices where supply equals demand in the market.

$$S_F(P_F) = D_F(P_F)$$

$$S_T(P_T) = D_T(P_T)$$

Solve these equations to find the equilibrium prices for $P_F$ and $P_T$.

**Application of Gross Substitutability**

Apply the principle of gross substitutability to the costs of spreading fake news $C_F$ and fact-checking cost $C_T$, and analyze how these costs influence market prices $P_F$ and $P_T$.

$$C_F = f(P_F, R_F, I_F)$$

$$C_T = g(P_T, R_T, I_T)$$

Here, $R_F, I_F, R_T, I_T$ are parameters representing the reliability and impact of fake news and fact-checking information, and $f$ and $g$ are cost functions that take these elements into account.

**Analysis of Interaction between Cost and Price**

Analyze the effects of the cost of spreading fake news $C_F$ and fact-checking cost $C_T$ on equilibrium prices $P_F$ and $P_T$, and reassess the equilibrium state of the information market. An increase in the cost of spreading fake news may lead to an increase in $P_F$ and a potential decrease in the supply of fake news. A decrease in fact-checking costs may lead to a decrease in $P_T$ and a potential increase in the supply of fact-checking information.

# 3. Previous Work

When considering the issue of the spread of fake news and the cost of fact-checking within the framework of gross substitutability (weak Walras' law), it is important to analyze the equilibrium of supply and demand in the information market. In this approach, the costs of spreading fake news and fact-checking function as "prices" in the information market and are believed to influence the behavior choices of news providers.

**Proposal of Formulas and Computational Process**

**Definition of Information Market**

Define the supply functions for fake news $F$ and fact-checking information $T$.

$$S_F(P_F) = a_F + b_F \cdot P_F$$

$$S_T(P_T) = a_T + b_T \cdot P_T$$

Here, $P_F$ and $P_T$ are the market prices for fake news and fact-checking information respectively, and $a_F, b_F, a_T, b_T$ are parameters.

Similarly, define the demand functions for fake news $F$ and fact-checking information $T$.

$$D_F(P_F) = c_F - d_F \cdot P_F$$

$$D_T(P_T) = c_T - d_T \cdot P_T$$

Here, $c_F, d_F, c_T, d_T$ are parameters.

**Calculation of Equilibrium Prices**

Calculate the equilibrium prices where supply equals demand in the market.

$$S_F(P_F) = D_F(P_F)$$

$$S_T(P_T) = D_T(P_T)$$

Solve these equations to find the equilibrium prices for $P_F$ and $P_T$.

**Application of Gross Substitutability**

Apply the principle of gross substitutability to the costs of spreading fake news $C_F$ and fact-checking cost $C_T$, and analyze how these costs influence market prices $P_F$ and $P_T$.

$$C_F = f(P_F, R_F, I_F)$$

$$C_T = g(P_T, R_T, I_T)$$

Here, $R_F, I_F, R_T, I_T$ are parameters representing the reliability and impact of fake news and fact-checking information, and $f$ and $g$ are cost functions that take these elements into account.

**Analysis of Interaction between Cost and Price**

Analyze the effects of the cost of spreading fake news $C_F$ and fact-checking cost $C_T$ on equilibrium prices $P_F$ and $P_T$, and reassess the equilibrium state of the information market.

An increase in the cost of spreading fake news may lead to an increase in $P_F$ and a potential decrease in the supply of fake news. A decrease in fact-checking costs may lead to a decrease in $P_T$ and a potential increase in the supply of fact-checking information.

Here we will introduce Metzler's concept into the previous discussions to consider the minimum cost path for detour routes. In the agent model aimed at resolving filter bubbles, by combining elements such as the Iterated Prisoner's Dilemma game, peer effects, and neighbor matrices, it is possible to design strategies to alleviate information bias and echo chamber phenomena. Incorporating concepts like rough substitutability (weak Walrasian law) and Metzler functions or Metzler matrices to consider the minimum cost path for detour routes is a methodological investigation into analyzing the complex dynamics of fake news diffusion and its punishment or compensation from a game theoretical perspective.

## 4. Discussion:Incorporation of rough substitutability (weak Walrasian law)

Rough substitutability is an economic principle stating that when the price of one good rises, demand for other goods increases. When applying this concept to agent models, if the "price" of fake news increases (i.e., it becomes costly due to punishment for its diffusion), agents are expected to increase their reliance on true information or trustworthy sources.

**Minimum cost path for detour routes and Metzler functions, Metzler matrices**

The problem of finding the minimum cost path for detour routes is crucial for optimizing the flow and diffusion patterns of information. Using Metzler functions or Metzler matrices, we can mathematically model the propagation paths of information and their costs. This enables the identification of diffusion paths for fake news within filter bubbles and the formulation of strategies to minimize them.

**Punishment-dominant problem and maximum compensation problem**

The punishment-dominant problem refers to the necessity for punishment against disseminators of fake news to exceed the gains from their dissemination. Conversely, the maximum compensation problem aims to provide sufficient incentives to agents contributing to the provision of reliable information or fact-checking. Analyzing these problems from a game theoretical perspective allows the design of strategies to prevent the diffusion of fake news and promote the sharing of truthful information.

**Analysis in game theory**

In the Iterated Prisoner's Dilemma game, agents update their strategies based on the history of past interactions. Strategies like the Tit-for-Tat strategy and the Grim Trigger strategy serve as criteria for determining how agents cooperate or defect towards each other. By applying these strategies and considering peer effects and neighbor matrices among agents, we can analyze how agents behave within filter bubbles and their impact on the diffusion of fake news or contributions to fact-checking.

Such a comprehensive approach enables the design of effective strategies to dissolve filter bubbles and maintain the integrity of information. Balancing punishment against disseminators of fake news and compensation for agents providing truthful information is essential for maintaining a healthy information environment.

When constructing agent models aimed at resolving the diffusion problem of fake news and resolving filter bubbles, considering the minimum cost path for detour routes using Metzler functions or Metzler matrices is an effective method for optimizing information propagation efficiency and suppressing the diffusion of fake news. Here, we propose modeling the problem of finding the minimum cost path for detour routes considering the resolution and diffusion of fake news, along with detailed formulas and computational processes.

**Minimum cost path for detour routes model**

Consider a network considering the strategic interactions between agents regarding the diffusion of fake news and the resolution of filter bubbles. In this network, each agent chooses whether to disseminate fake news, provide truthful information, or remain indifferent. The interactions between agents are represented using Metzler matrices, and costs and gains for each agent's action selection are defined.

**Metzler functions and Metzler matrices**

A Metzler matrix is a real matrix with non-negative off-diagonal elements. Using this matrix to represent interactions between agents, we can model information propagation patterns and their costs mathematically. The Metzler function calculates the expected gains for an agent based on its action selection.

**Expected gain function**

The expected gain $E[U_i]$ for agent $i$ based on its action selection is expressed as follows:

$$E[U_i] = \sum_{j \neq i} a_{ij} \cdot (U_{ij} C_{ij})$$

Here, $a_{ij}$ represents the influence of information propagation from agent $i$ to agent $j$ in the Metzler matrix, $U_{ij}$ represents the gain for agent $i$ from agent $j$ accepting its information, and $C_{ij}$ represents the cost for agent $i$ to propagate information to agent $j$.

**Idea of computational process**

Building the network model Define the Metzler matrix representing interactions between agents. Each element $a_{ij}$ of the matrix indicates the influence of information propagation from agent $i$ to agent $j$.

Definition of costs and gains Define the cost $C_{ij}$ and gain $U_{ij}$ for each agent's action selection. It is common to set high costs and low gains for fake news dissemination and low costs and high gains for providing truthful information. Calculation of expected gains Calculate the expected gains $E[U_i]$ for each agent using the above expected gain function. This expected gain serves as a criterion for agents' action selection. Strategy update Agents update their action selection to maximize their expected gains. If the cost for disseminating fake news is high, agents tend to choose actions that provide truthful information. Using this model, it is possible to design strategies to suppress the diffusion of fake news and promote the provision of truthful information. Furthermore, it is expected to contribute to improving the behavior of agents towards resolving filter bubbles.

When considering the minimum cost path model for detour routes in the context of fake news and fact-checking, it is crucial to construct a network taking into account the cost and effectiveness of information transmission between each news provider (agent). Below, we outline the detailed equations and computational process for this.

**Definition of the Network Model**

The network models news providers as nodes and models their interactions (information transmission) as edges. The weight of each edge reflects the cost of information transmission.

Node set $N = \{n_1, n_2, \ldots, n_k\}$
Edge set $E = \{(n_i, n_j) | n_i, n_j \in N\}$
Cost function $C : E \to \mathbb{R}^+$

**Construction of the Cost Matrix**

Construct the cost matrix $M$ of the network. Each element $M_{ij}$ of the matrix represents the cost of information transmission from node $n_i$ to node $n_j$. If information transmission is impossible, the cost is set to infinity ($\infty$).

**Definition of the Metzler Matrix**

Adjust the cost matrix $M$ to positive values to define the Metzler matrix, as it requires non-negative off-diagonal elements.

**Calculation of the Minimum Cost Path**

Use algorithms such as Dijkstra's algorithm or the Bellman-Ford algorithm to calculate the minimum cost path between any two points. This identifies the most efficient transmission paths for fake news or fact-checking information.

**Identification of Detour Routes**

Calculate routes bypassing specific nodes (sources of fake news or centers of fact-checking) to avoid the spread of fake news or identify effective transmission routes for fact-checking information. This is done by excluding specific nodes and recalculating the minimum cost path.

**Idea of the Calculation**

Let $P_{ij}$ denote the minimum cost path from node $n_i$ to node $n_j$, then the minimum cost $C_{ij}$ is calculated as follows:

$$C_{ij} = \min_{P_{ij}} \sum_{(n_a,n_b) \in P_{ij}} M_{ab}$$

Here, $P_{ij}$ represents the path from $n_i$ to $n_j$, and $M_{ab}$ is the cost of information transmission from node $n_a$ to node $n_b$.

By using this model, it is possible to identify detour routes that minimize the spread of fake news or optimal routes to spread fact-checking information maximally. Additionally, incorporating information from a wider range of sources, rather than relying solely on information within filter bubbles, is expected to enhance the diversity and quality of information.

When modeling the cost of actions towards informational health and malicious dissemination using Metzler functions and Metzler matrices, we expand detailed equations and computational processes following the steps outlined below:

**Definition of the Network Model**

Node set $N = \{n_1, n_2, \ldots, n_k\}$. Each node represents a news provider. Edge set $E = \{(n_i, n_j) | n_i, n_j \in N \text{ and } i \neq j\}$. Edges represent information transmission between news providers.

**Definition of the Cost Function**

Define the cost function for information transmission $C : E \to \mathbb{R}^+$. This function considers the cost of both positive and malicious information transmission.

Positive information transmission cost $C_{ij}^+$ is the cost of transmitting positive information (e.g., fact-checking information) from node $n_i$ to node $n_j$. Malicious information transmission cost $C_{ij}^-$ is the cost of transmitting malicious information (e.g., fake news) from node $n_i$ to node $n_j$.

**Construction of the Metzler Matrix**

The Metzler matrix $M$ is a matrix with non-negative real off-diagonal elements. Define $M$ as the cost matrix for information transmission, and $M_{ij}$ as the total cost of transmitting information from $n_i$ to $n_j$.

$$M_{ij} = \begin{cases} C_{ij}^+ & \text{if positive information transmission} \\ C_{ij}^- & \text{if malicious information transmission} \\ 0 & \text{if } i = j \\ \infty & \text{if transmission is impossible} \end{cases}$$

**Calculation of the Minimum Cost Path**

Calculate the minimum cost path based on the Metzler matrix $M$ using algorithms such as Dijkstra's algorithm or the Bellman-Ford algorithm. This identifies information transmission routes to improve informational health and paths to prevent the spread of malicious information.

**Example Equation**

The cost function for finding the minimum cost path from node $n_i$ to $n_j$ is given by:

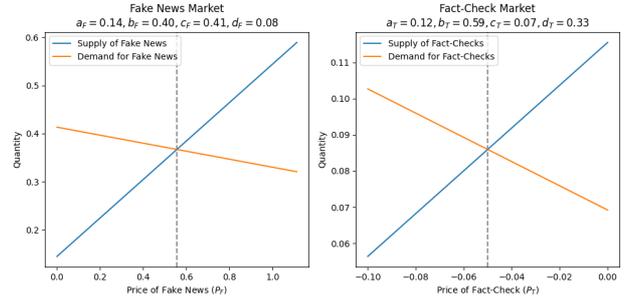

Fig. 3: Price of Fact-Check, Price of Fake News

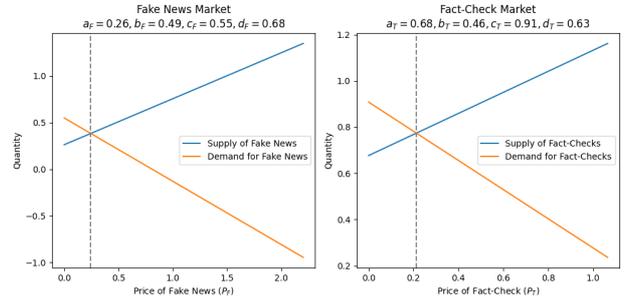

Fig. 4: Price of Fact-Check, Price of Fake News

$$C_{min}(i,j) = \min \left( \sum_{(n_a,n_b) \in P_{ij}} M_{ab} \right)$$

Here, $P_{ij}$ represents the path from $n_i$ to $n_j$, and $M_{ab}$ is the cost of information transmission from node $n_a$ to node $n_b$.

By using this model, strategies can be developed to promote positive information transmission contributing to informational health and to suppress the dissemination of malicious information such as fake news. Considering specific cost settings and the effectiveness of information transmission allows for a more realistic modeling.

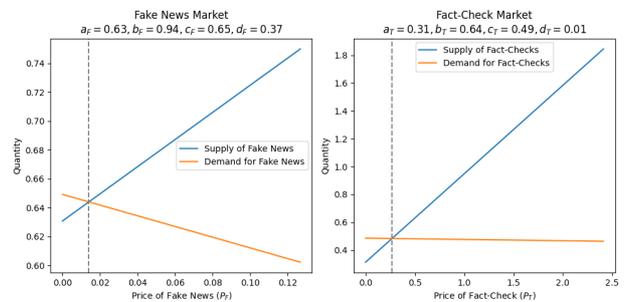

Fig. 5: Price of Fact-Check, Price of Fake News

**Analysis Based on Provided Equations**

**Defining the Information Market**

Supply functions are given as $S_F(P_F) = a_F + b_F \cdot P_F$ for fake news and $S_T(P_T) = a_T + b_T \cdot P_T$ for fact-checks. Demand functions are given as $D_F(P_F) = c_F - d_F \cdot P_F$ for fake news and $D_T(P_T) = c_T - d_T \cdot P_T$ for fact-checks.

In each graph, the blue line represents the supply function, and the orange line represents the demand function for their respective markets.

**Calculating Equilibrium Prices**

To find the equilibrium prices, we would set the supply function equal to the demand function for each market:

$a_F + b_F \cdot P_F = c_F - d_F \cdot P_F$ for the fake news market.
$a_T + b_T \cdot P_T = c_T - d_T \cdot P_T$ for the fact-check market.

The equilibrium price is where the supply and demand curves intersect in each graph.

**Applying Gross Substitutability**

The principle of gross substitutability in this context would consider how the costs of spreading fake news and the costs of fact-checking influence each other. These costs are represented in the graphs by the price axis for each market.

**Analyzing the Interplay of Costs and Prices**

An increase in the cost of spreading fake news would shift the supply curve for fake news leftward, increasing $P_F$. A decrease in the cost of fact-checking would shift the supply curve for fact-checks rightward, decreasing $P_T$.

**Assessing the Impact on the Health of the Information Market**

By analyzing the changes in costs, one can evaluate the impact on the market's health and propose strategies to suppress the spread of fake news and promote fact-checking.

**Specific Graph Analysis**

In the first Results, the graphs show the following parameters for the fake news market and fact-check market, respectively:

$a_F = 0.63, b_F = 0.94, c_F = 0.65, d_F = 0.37$ $a_T = 0.31, b_T = 0.64, c_T = 0.49, d_T = 0.01$

The equilibrium prices are not given directly, but they are where the supply and demand curves intersect on each graph. The dashed vertical lines likely represent the equilibrium price levels.

Results represent the supply and demand dynamics in a market, specifically for "fake news" and "fact-checks." In the context of minimizing the cost of spreading fake news and maximizing the compensation for fact-checking.

**Fake News Market**

Supply and Demand Relationship, In this market, as the price of fake news increases, the quantity supplied also increases, indicating a direct relationship. Conversely, the quantity demanded decreases, showing an inverse relationship between price and demand.

**Equilibrium Price**

The point where the supply and demand curves intersect represents the market equilibrium, where the quantity supplied equals the quantity demanded. This price signals an efficient market state where no excess supply or demand exists.

**Fact-Check Market**

Supply and Demand Relationship, Similar to the fake news market, as the price of fact-checks increases, the quantity supplied increases. However, the demand curve shows that the quantity demanded also increases with the price, which could indicate a perceived value in fact-checks that increases with cost or possibly an inelastic demand.

**Negative Prices**

The fact-check market graph shows a portion where the price is negative, which is unusual in a typical market scenario. This might represent a situation where the cost of not performing fact-checks (i.e., the negative consequences of spreading fake news) is considered, hence the negative price.

**Minimizing Cost Routes in Fake News Spread**

In a network context, minimizing cost routes could refer to finding the least costly paths for spreading information. For fake news, this would be the paths that allow the news to spread most effectively at the lowest cost. This is analogous to minimizing the "price" in the graph, where cost routes are reduced to increase the spread. In a network graph, these would be the shortest or most direct paths between nodes.

**Maximizing Compensation in Fact-Check Spread**

Conversely, the maximum compensation problem in the context of fact-checking could refer to maximizing rewards for spreading true information. In the graph, this would involve increasing the "price" of fact-checks to maximize the quantity supplied, incentivizing more agents to engage in fact-checking. This scenario would favor strategies that amplify the dissemination of factual information, thereby promoting information health.

In summary, the approach to these issues would involve adjusting network strategies to affect the spread of information, analogous to altering prices to influence supply and demand in traditional markets. This could be done by increasing the cost of spreading fake news (thus making it less attractive) and increasing the incentives for spreading fact-checks (making it more attractive).

## 5. Discussion:Cost Matrix

In regard to constructing the cost matrix, detailed equations and computational processes are shown below.

### Definition of the Cost Matrix

The cost matrix $C$ represents the cost of information transmission between news providers. The size of the matrix is determined by the number of news providers (nodes), resulting in an $n \times n$ matrix.

### Setting the Cost Function

Define a function $C(i, j)$ to represent the cost of transmitting information from each news provider $i$ to $j$. This cost varies depending on the nature of the information (positive or malicious).

The cost of positive information transmission $C^+(i, j)$ is the cost of transmitting positive information (e.g., fact-checking information) from news provider $i$ to $j$.

The cost of malicious information transmission $C^-(i, j)$ is the cost of transmitting malicious information (e.g., fake news) from news provider $i$ to $j$.

### Concrete Formulation of Equations

The cost function $C(i, j)$ is concretely formulated as follows:

$$C(i,j) = \begin{cases} C^+(i,j) & \text{if positive information transmission} \\ C^-(i,j) & \text{if malicious information transmission} \\ \infty & \text{if transmission is impossible} \end{cases}$$

### Construction of the Cost Matrix $C$

Set each element $C_{ij}$ of the cost matrix $C$ based on the above cost function $C(i, j)$. This reflects all information transmission costs between news providers in the matrix.

$$C = \begin{bmatrix} C(1,1) & C(1,2) & \cdots & C(1,n) \\ C(2,1) & C(2,2) & \cdots & C(2,n) \\ \vdots & \vdots & \ddots & \vdots \\ C(n,1) & C(n,2) & \cdots & C(n,n) \end{bmatrix}$$

### Example Calculation Process

For example, suppose there are three news providers $A, B, C$. If the cost of positive information transmission from $A$ to $B$ is 2, the cost of malicious information transmission from $B$ to $C$ is 5, and direct transmission from $A$ to $C$ is impossible, the cost matrix $C$ would be as follows:

$$C = \begin{bmatrix} 0 & 2 & \infty \\ \infty & 0 & 5 \\ \infty & \infty & 0 \end{bmatrix}$$

Using this matrix, we can identify the optimal paths for transmitting information at minimum cost. Through this process, strategies can be formulated to improve informational health and to prevent the spread of fake news.

## 6. Discussion:Punishment-dominant Problem

The concepts of the punishment-dominant problem and the maximum compensation problem are important in game theory and economics for analyzing the incentive structure of participants' strategic choices. In the context of fake news and fact-checking, these concepts can be used to understand and model the interactions between news providers and the strategic impact of their outcomes.

**Punishment-dominant Problem**

The punishment-dominant problem refers to situations where a player adopts punitive strategies to deter other players from non-cooperative behavior. In the context of fake news, this corresponds to punitive responses to spreaders of fake news or providing incentives to correct misinformation.

**Equations and Computational Process**

1. Definition of the Punishment-dominant Function

$$P(i,j) = \begin{cases} l & \text{if } s_j = D \text{ and } s_i = C \\ 0 & \text{otherwise} \end{cases}$$

Here, $P(i, j)$ represents the magnitude of punishment player $i$ applies to player $j$, $s_j$ is player $j$'s strategy (cooperation $C$ or defection $D$), and $l$ is the level of punishment.

2. Evaluation of Punishment-dominant Impact Compute the total punishment for each player and reflect its impact on strategic choices.

$$L_i = \sum_{j \neq i} P(i,j)$$

Here, $L_i$ is the total punishment for player $i$.

**Maximum Compensation Problem**

The maximum compensation problem refers to achieving the maximum benefit by players through cooperative actions. In the context of fake news, this includes rewards for sharing accurate information or incentives for participating in fact-checking.

**Equations and Computational Process**

1. Definition of the Compensation Function

$$C(i,j) = \begin{cases} m & \text{if } s_i = C \text{ and } s_j = C \\ 0 & \text{otherwise} \end{cases}$$

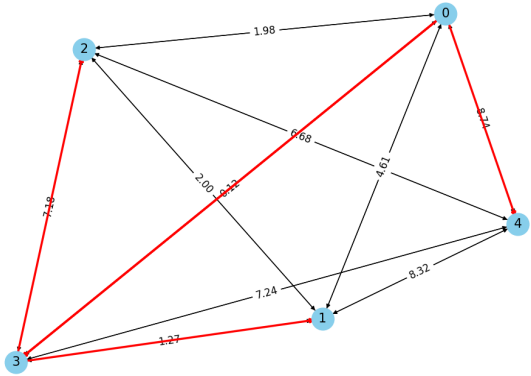

Fig. 6: Minimum Cost Paths from Node

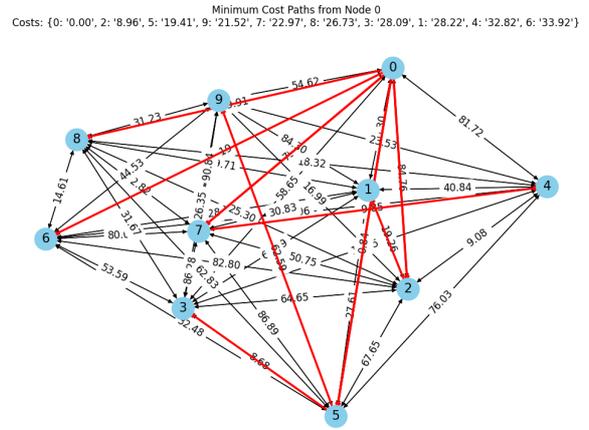

Fig. 8: Minimum Cost Paths from Node

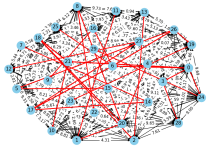

Fig. 7: Minimum Cost Paths from Node

Here, $C(i, j)$ represents the compensation player $i$ obtains by cooperating with player $j$, and $m$ is the level of compensation.

2. Evaluation of Maximum Compensation Impact Compute the total compensation for each player and reflect its impact on strategic choices.

$$M_i = \sum_{j \neq i} C(i, j)$$

Here, $M_i$ is the total compensation for player $i$.

Using these equations and computational processes, it's possible to model the strategic interactions between news providers and evaluate the impact on informational health through the balance of punishment-dominant and maximum compensation problems. These concepts play a crucial role in designing strategic incentives to suppress the spread of fake news and promote participation in fact-checking.

Network models with nodes representing news providers and edges representing information transmission paths. The red lines depict the minimum cost paths from a specific node (presumably node 0 in these cases) to all other nodes in the network.

**Metzler Matrix Construction (Step 1)**

A Metzler matrix is a matrix where all off-diagonal elements are non-negative. In this context, it is used to represent the total cost of information transmission between nodes.

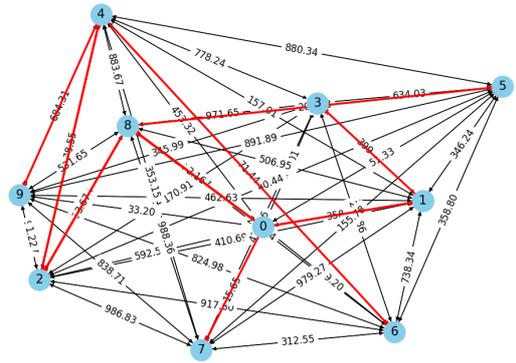

Fig. 9: Minimum Cost Paths from Node

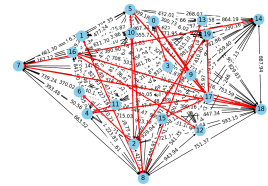

Fig. 10: Minimum Cost Paths from Node

The value $M_{ij}$ is the cost from node $n_i$ to $n_j$, which could either be the cost of transmitting positive information ($C_{ij}^+$) or malicious information ($C_{ij}^-$), or it could be zero (if $i = j$) or infinity (if transmission is not possible).

**Calculating Minimum Cost Paths (Step 2)**

The minimum cost path function $C_{min}(i, j)$ calculates the least costly path for information transmission from node $n_i$ to $n_j$ by summing the costs $M_{ab}$ along the path $P_{ij}$ from node $n_a$ to $n_b$.

**Fig.6**

A simple network with 5 nodes where the red lines indicate the minimum cost paths from node 0. The costs associated with these paths are also given, suggesting the model has been applied to find the most efficient way to disseminate information from node 0.

**Fig.7**

Fig. 7 much more complex network with many nodes. Again, the red lines highlight the most efficient paths from node 0 to all other nodes. The cost structure is more intricate due to the greater number of nodes and potential paths.

**Fig.8**

An even larger network, suggesting a complex system of information providers where the strategy for transmitting information or preventing malicious spread would be quite comprehensive due to the number of involved parties.

These results exemplify how a Metzler matrix and associated minimum cost path analysis can be used to strategize the dissemination of truthful information and suppress the spread of fake news in a network. Such models can guide interventions to reinforce positive information flow and identify critical nodes or paths that require monitoring to prevent the spread of harmful information.

The results network graphs with nodes and edges that represent the framework for modeling the flow of information and the associated costs of spreading both genuine and malicious information. These graphs are used to determine the minimum cost paths from a given source node to all other nodes within the network, which is critical for understanding how information propagates and where interventions may be necessary to ensure the healthy spread of information.

Results represent network graphs with minimum cost paths highlighted, along with the given parameters (like weights w1, w2, w3), are visual representations of networks where each node could represent a news provider or user, and the edges represent the flow of information between them.

The costs associated with each edge could represent the difficulty or "cost" of transmitting information from one node to another. In the context of fake news spread and containment, these graphs can be interpreted in light of the Metzler function and the concepts of punishment-dominant and maximum compensation problems.

**Minimum Cost Paths and Fake News Spread**

The minimum cost paths in these networks can be considered as the most efficient routes for information to travel through the network. From the perspective of fake news spread: Lower costs (lighter weights) on edges could represent easy transmission of fake news, due to factors like shared ideologies, lack of fact-checking, or high trust between nodes. Higher costs (heavier weights) could represent barriers to the spread of fake news, which might be due to skepticism, better fact-checking, or a lower tendency to share unverified information.

When considering the spread of fake news, one would look for routes in the network that have lower costs, as these are the paths through which fake news could spread more rapidly and widely.

**Metzler Function Application**

The Metzler function is used to calculate the costs of interactions between nodes, taking into account the truthfulness, importance, and urgency of information. The weights $w_1, w_2, w_3$ could represent the importance of these factors in determining the cost of information transmission.

A high weight on truthfulness ($w_1$) might indicate that the network strongly penalizes the spread of fake news, thereby increasing the cost of transmission between nodes if the news is fake. - A higher weight on urgency ($w_3$) could mean that the cost to transmit urgent information is less, facilitating its spread in the network.

**Punishment-dominant and Maximum Compensation Problems**

**From a game-theoretic standpoint**

Punishment-dominant strategies might be represented in the network by high costs for transmitting fake news, deterring nodes from sharing it. Maximum compensation strategies could be represented by lower costs for transmitting truthful, important, and urgent news, incentivizing nodes to spread such information.

**Strategic Implications**

The network analysis can be used to devise strategies for containing the spread of fake news by identifying key nodes or edges where interventions could be most effective. Increasing the costs of certain connections could represent introducing fact-checking measures or penalizing the spread of fake news. Reducing costs for certain paths could involve incentivizing the sharing of verified information or improving the literacy of news consumers to discern truth from falsehood.

The Floyd-Warshall algorithm mentioned in your description is a computational process used to find the shortest paths in a weighted network. In this context, it would help identify the minimum cost paths for the spread of information, which, depending on the strategy, could either be the paths to target for containment or for the promotion of accurate information dissemination.

In summary, by examining the network's structure, cost dynamics, and the minimum cost paths, one can gain insights into the spread and containment of fake news. The strategic application of the Metzler function to increase or decrease costs can inform policies or measures to promote informational health and combat misinformation.

**Analysis Based on the Steps Provided**

**Network Model Definition**

Nodes ($N$) represent news providers within the network. Edges ($E$) represent the potential paths for information transfer between the nodes.

**Cost Function Definition**

Costs for the transmission of information are defined for both positive ($C_{ij}^+$) and malicious ($C_{ij}^-$) information transfer between nodes $n_i$ and $n_j$.

**Metzler Matrix Construction**

The Metzler matrix ($M$) is created with non-negative off-diagonal elements to represent the total costs of information transmission.

**Minimum Cost Path Calculation**

Algorithms like Dijkstra's or Bellman-Ford are used to compute the minimum cost paths based on the Metzler matrix ($M$), identifying the most efficient routes for the spread of beneficial information and containment of malicious information.

In the context of the graphs (Fig.9 and Fig.10), the red lines represent the minimum cost paths from node 0 to all other nodes, calculated using the principles mentioned above. Each edge on these paths is labeled with a cost, which likely corresponds to the values in the Metzler matrix ($M$).

The costs noted at the top of each image correspond to the minimum total cost from the source node (node 0) to each other node. These values are critical for strategizing interventions in the network. For example: Nodes with lower costs to reach might be more influential due to their easier accessibility for information spread. Nodes with higher costs may represent bottlenecks or challenging areas where spreading either genuine or malicious information is more difficult.

By analyzing these costs and paths, strategies can be developed to enhance the spread of factual information and to mitigate the dissemination of fake news. This involves identifying key nodes that can be targeted for the distribution of fact-checks or monitoring to prevent the spread of misinformation.

In summary, these graphs are a visual tool to help strategize the propagation of information in a network by calculating and analyzing the costs associated with different transmission paths. They can be used to inform decisions on where to focus efforts for spreading factual information and combating misinformation effectively.

# 7. Discussion:Walras' law

The concept of informational substitutability, commonly known as Walras' law, is one of the principles concerning market supply-demand equilibrium in economics. This law states that if the supply and demand for all goods in one market are in equilibrium, then the supply and demand will also be in equilibrium in all other markets. In the context of informational health, this principle can be applied to analyze the balance of information supply and demand in the "market" of fake news and fact-checking information.

**Theoretical Supplement**

To apply the concept of informational substitutability to the context of fake news and fact-checking, it is necessary to quantitatively evaluate the "value" and "cost" of information. The value of information varies based on factors such as its reliability, accuracy, and influence, while the cost of information is determined by the effort and resources required to obtain, verify, and share it.

**Equations and Computational Process**

(1) **Information Value Function**

$$V(i) = \alpha R(i) + \beta A(i)$$

Here, $V(i)$ is the value of information $i$, $R(i)$ is the reliability of information $i$, $A(i)$ is the influence of information $i$, and $\alpha$ and $\beta$ are coefficients.

(2) **Information Cost Function**

$$C(i) = \gamma T(i) + \delta S(i)$$

Here, $C(i)$ is the cost of information $i$, $T(i)$ is the time required to obtain information $i$, $S(i)$ is the effort required to share it, and $\gamma$ and $\delta$ are coefficients.

(3) **Supply-Demand Equilibrium Condition** The condition for supply-demand equilibrium of information in

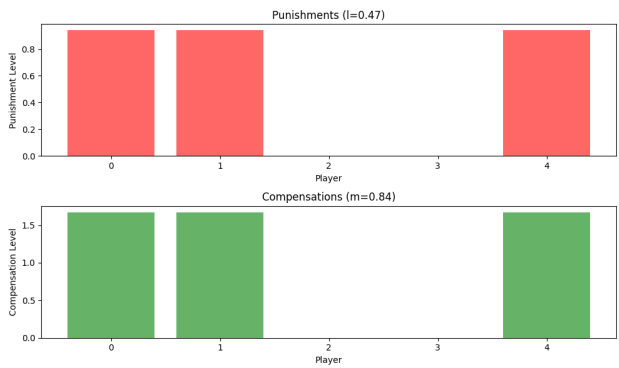

Fig. 11: Compensation Level, Punishment Level

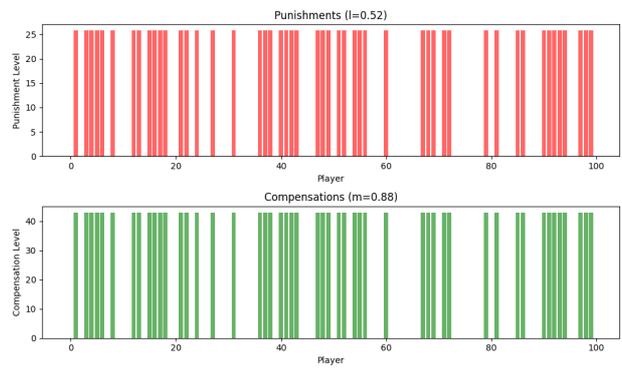

Fig. 13: Compensation Level, Punishment Level

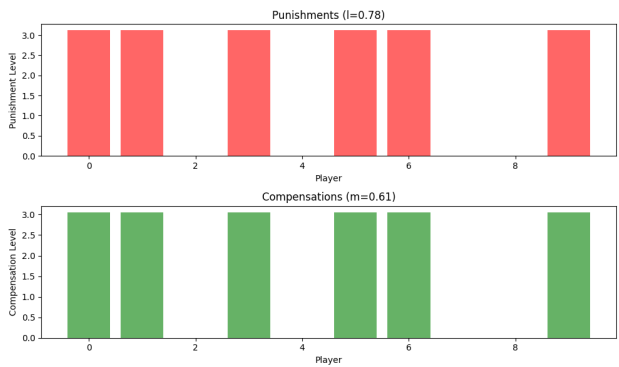

Fig. 12: Compensation Level, Punishment Level

the market is when the value and cost of information are equal.

$$V(i) = C(i)$$

When this condition is met, the supply and demand of information are in equilibrium.

(4) **Application of Walras' Law** In the market of fake news and fact-checking information, if the above supply-demand equilibrium condition holds for all information, then the overall information market is in equilibrium.

**Application**

Using this model, strategies can be devised to prevent the spread of fake news and promote the dissemination of fact-checking information. For example, by increasing the value of fact-checking information, it is possible to change its equilibrium point of supply and demand, thereby encouraging the supply of more fact-checking information. Additionally, by increasing the cost of fake news, its supply can be reduced.

The use of the concept of informational substitutability enables an understanding of the balance of supply and demand in the information market and the development of effective approaches to improving informational health.

The images you've provided, labeled Fig.10-12, contain bar graphs that depict punishment levels and compensation levels applied to different players in a strategic interaction, likely within the context of a game theoretical model related to the spread of information, such as fake news and fact-checking.

**Punishment-dominant Problem**

In these graphs, the punishment-dominant problem is represented by the first set of bar charts (red bars), with the level of punishment that each player incurs. The level of punishment $l$ is applied by a player $i$ to a player $j$ if player $j$ chooses to defect (spread fake news, in this context) and player $i$ cooperates (performs fact-checking or promotes truthful information). The equation provided calculates the total level of punishment each player faces based on their interactions with all other players in the game.

**Maximum Compensation Problem**

The second set of bar charts (green bars) represents the maximum compensation problem. Here, compensation is given to players that cooperate with each other (share truthful information or participate in fact-checking). The equation calculates the total compensation a player receives based on cooperative interactions with all other players.

The bar charts visually represent the level of punishment and compensation for each player in the game. The values of $l$ and $m$ mentioned in the graphs (for punishment and compensation respectively) indicate the intensities of these strategies. In the context of fake news, punishment could represent fines or other penalties for spreading misinformation, while compensation could represent rewards or incentives for sharing accurate information or fact-checking. High punishment levels indicate a strategy where deterring misinformation by punitive means is emphasized. High compensation levels suggest a strategy that focuses on encouraging cooperative behavior, such as sharing verified information.

The players in the graphs are likely to be news providers or individuals in a social network. The patterns in the graphs indicate how each player is being affected by the punitive and compensatory strategies implemented within the game.

**Strategic Implications**

By analyzing these bar graphs, one can infer the overall strategy of the system: If punishments are generally high, it indicates a strict regime against the spread of fake news. If compensations are generally high, it indicates strong incentives for spreading truthful information.

This analysis can help policymakers or platform designers to understand the potential effectiveness of different incentive structures for encouraging truthful information sharing and discouraging the spread of fake news. By adjusting the levels of punishment and compensation, they can potentially influence the behavior of players in the network towards a more desired outcome.

Resuls, we have data on punishments and compensations assigned to various players, which can be related to the dynamics of spreading fake news and the incentives for spreading true information.

**Punishments**

The punishment bars appear to represent the punitive measures taken against players who may have engaged in spreading fake news or uncooperative behavior. From a game-theoretic perspective, the concept of punishment-dominant strategies is applied to deter players from engaging in negative actions such as spreading misinformation.

High Punishment Levels (l=0.52), If we look at the high levels of punishment across players, it suggests that the model heavily penalizes the spread of fake news. The uniformity of punishment levels across all players can indicate a strategy where every player is held to the same standard of truthfulness.

Variable Punishment Levels (l=0.78, l=0.47), Different levels of punishment could imply that different players have different roles or influence within the network. Players with higher punishment levels might be key spreaders of information, and deterring them is crucial.

The compensation bars seem to represent the rewards given to players for cooperative actions, such as spreading true information or engaging in fact-checking activities. High Compensation Levels (m=0.88), A high compensation level across players would encourage the spread of verified information. Uniform high compensation can incentivize a collective effort towards maintaining informational integrity. Variable Compensation Levels (m=0.61, m=0.84), This could reflect the varying effectiveness or credibility of players in spreading true information. Players with higher compensation levels might be more trusted or effective in fact-checking efforts.

The punishment and compensation models offer a way to manage the strategic behavior of players in the context of information dissemination: For Fake News Spread Risk, Punishment acts as a deterrent against the spread of misinformation. By increasing the cost of spreading fake news through punitive measures, the network can reduce the spread of such news. For Maximum Compensation Issues, Compensation encourages the propagation of true information. By rewarding players for sharing accurate content and engaging in fact-checking, the network promotes informational health. Balancing Punishments and Compensations, To effectively manage fake news and encourage the spread of true information, there must be a balance between punitive measures for misinformation and incentives for accurate reporting. Targeted Interventions, Understanding which players hold significant influence in the network allows for targeted interventions, where key nodes can be specifically deterred from spreading fake news or rewarded for disseminating true information.

In essence, these measures are about modifying the cost-benefit analysis of each player within the network to align their incentives with the goal of enhancing the quality of information flow, thereby improving the overall informational ecosystem.

## 8. Conclusion: Weak Walras' law, in agent-based models to resolve filter bubbles in the repeated prisoner's dilemma game

The process of applying the concept of informational substitutability, namely the weak Walras' law, in agent-based models to resolve filter bubbles in the repeated prisoner's dilemma game involves constructing equations and computational steps as follows:

**Definition of Agent Actions and Payoffs** Define the actions available to agents as cooperation $C$ and non-cooperation $D$, and define the payoff for each action:

$$U(C,C) > U(D,C) > U(D,D) > U(C,D)$$

Here, $U(x, y)$ represents the payoff obtained by an agent taking action $x$ when the opponent takes action $y$.

**Modeling of Information Value and Cost** Model the value $V(i)$ and cost $C(i)$ of information $i$ within the filter bubble for each agent:

$$V(i) = \alpha R(i) + \beta A(i)$$

$$C(i) = \gamma T(i) + \delta S(i)$$

Where $R(i)$ represents the reliability of information $i$, $A(i)$ represents the influence, $T(i)$ represents the time required for acquisition, and $S(i)$ represents the effort required for sharing.

**Application of Supply-Demand Equilibrium and Weak Walras' Law** Construct a model where each agent selects actions based on the value and cost of information, leading to supply-demand equilibrium within the information market:

$$V(i) = C(i)$$

If this equilibrium condition holds for all information, the entire market is considered to be in an equilibrium state.

**Strategies for Resolving Filter Bubbles** Introduce strategies to enhance the value (increase $\alpha$ and $\beta$) or reduce the cost (decrease $\gamma$ and $\delta$) of information to resolve filter bubbles, and analyze their effects:

$$V'(i) = \alpha' R(i) + \beta' A(i)$$

$$C'(i) = \gamma' T(i) + \delta' S(i)$$

Where $\alpha', \beta', \gamma', \delta'$ are the new coefficients after strategy implementation.

**Analysis of Strategy Effects** Use the new value $V'(i)$ and cost $C'(i)$ to determine the new supply-demand equilibrium in the information market and evaluate the effectiveness in resolving filter bubbles:

$$V'(i) = C'(i)$$

**Computational Example**

For a specific computational example, define the initial value and cost of information circulating within a filter bubble, simulate how the proposed strategy affects the value and cost, and ultimately how it changes the supply-demand equilibrium.

This approach provides a foundation for understanding the mechanism of information circulation within filter bubbles and devising effective strategies to improve informational health.

For detailed equations and computational steps in Steps 2 and 4, they are as follows:

**Definition of the Metzler Function**

Using the Metzler function, define the cost of information propagation between nodes. This cost varies based on factors such as the truthfulness, importance, and urgency of information. A specific functional form could be set as follows:

$$C_{ij} = w_1 \times T + w_2 \times I + w_3 \times U$$

Where: $C_{ij}$ is the cost of information propagation from node $i$ to node $j$. $T$ represents the truthfulness of information (higher cost for fake news, lower cost for true news). $I$ represents the importance of information (higher cost for important information). $U$ represents the urgency of information (higher cost for urgent information). $w_1, w_2, w_3$ are the weights for each factor.

**Calculation of Minimum Cost Paths**

To compute the minimum cost paths between all nodes, commonly used algorithms such as Floyd-Warshall algorithm are employed. This algorithm efficiently finds the shortest paths between all pairs of nodes.

The main steps of the Floyd-Warshall algorithm are as follows: Initialization of the cost matrix $M$: Each element $M_{ij}$ of $M$ is set to the cost $C_{ij}$ of the edge $(i, j)$. If the edge does not exist, the cost is set to infinity (or a very large value). For each node $k$, update the cost for all pairs $(i, j)$ of nodes:

$$M_{ij} = \min(M_{ij}, M_{ik} + M_{kj})$$

Where $M_{ij}$ is the minimum cost of a direct or indirect path from node $i$ to node $j$.

By executing this algorithm, the minimum cost paths between all nodes are stored in $M$. Finally, this information can be used to evaluate strategies for minimizing the spread of fake news or identifying effective dissemination routes for fake news.

The computational process when applying informational substitutability, namely weak Walras' law, is akin to finding the balance of supply and demand in a market. In this case, the "market" represents the environment for information circulation, and the "goods" are the information provided by news providers (both true and fake news). From the perspective of informational substitutability, the preferences of agents (news providers) are continuous, and their behavioral choices affect the equilibrium of the information market.

1. **Preferences and Utility Functions:** Each news provider $i$ has preferences regarding sharing information, represented by the utility function $U_i$. This function is often modeled as a function of the truthfulness $T$, importance $I$, and urgency $U$ of information.

2. **Supply and Demand Functions:** The supply function $S_i$ indicates how much information the news provider $i$ supplies to the market. This depends on the provider's cost structure and external incentives. The demand function $D_j$ indicates how much information other news providers $j$ consume (accept, disseminate). This depends on the quality of the provided information and the preferences of the recipients.

3. **Definition of Equilibrium:** Market equilibrium in the information market occurs when the supply and demand of all news providers match. In other words, for all $i$, $S_i = D_i$ holds.

4. **Computing Equilibrium:** To find equilibrium, solve the supply and demand functions as a system of simultaneous equations. Numerical analysis methods such as Newton's method or fixed-point algorithms may be used.

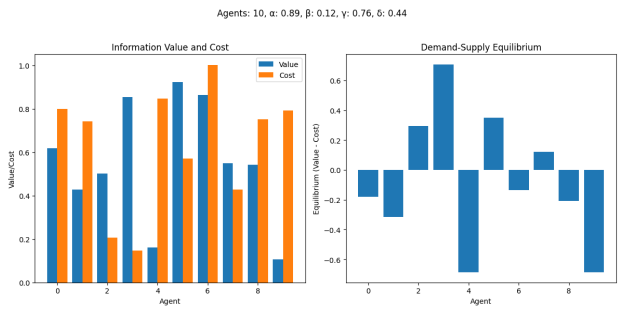

Fig. 14: Information Value and Cost / Demand-Supply Equilibrium

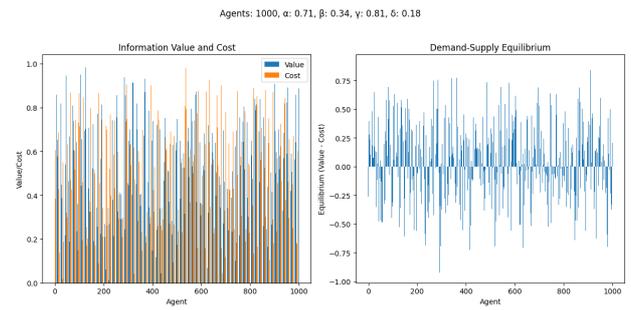

Fig. 16: Information Value and Cost / Demand-Supply Equilibrium

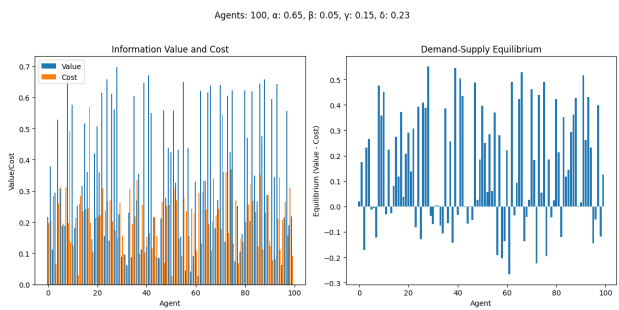

Fig. 15: Information Value and Cost / Demand-Supply Equilibrium

**Example Equations** As examples of supply and demand functions, consider the following linear forms:

Supply function: $S_i = a_i + b_i \cdot P$ Demand function: $D_i = c_i d_i \cdot P$

Here, $a_i, b_i, c_i, d_i$ are parameters, and $P$ represents the "price" of information. This price metaphorically represents the quality and reliability of information and is influenced by the cost and utility for news providers to provide information.

Setting up these functions for all news providers and finding $P$ where $S_i = D_i$ for all $i$ allows for the computation of market equilibrium. This equilibrium indicates the optimal circulation state of information in the context of fake news and fact-checking, providing a basis for evaluating its impact on informational health.

Results represent a model of agent behavior in a market-like setting, specifically related to the exchange of information. These graphs are outputs from a simulation or a model designed to explore how agents value information, how much it costs them, and how these factors reach an equilibrium in a system that represents the spread of information, such as news.

**Information Value and Cost**

The first part of each pair of graphs, titled "Information Value and Cost," presents a comparison of the value and cost of information for each agent. We can see that: The blue bars represent the value of the information to each agent. The orange bars represent the cost of acquiring or sharing that information. The parameters $\alpha$, $\beta$, $\gamma$, and $\delta$ mentioned in the titles are likely to be coefficients that determine how the value and cost are calculated based on factors like reliability, impact, time, and effort associated with the information.

**Demand-Supply Equilibrium**

The second part of each pair of graphs, titled "Demand-Supply Equilibrium," seems to show the equilibrium state for each agent, calculated as the difference between the value and the cost of the information. A positive value indicates that the value of information exceeds its cost, while a negative value suggests that the cost is greater than the value to the agent.

**This model might be used to analyze situations such as: How do agents react to different types of information?**

What is the cost threshold above which agents are no longer willing to acquire or share information? How does the overall market for information reach an equilibrium based on individual agent behaviors?

**Analysis Based on the Uploaded Graphs**

**From the graphs, we can infer several things**

Agents are not uniform in how they value or what they pay for information. There is a variation in both the value and the cost among agents, which suggests diverse behaviors or preferences. The equilibrium state varies widely among agents, indicating that some find greater value in the information than others, or that the costs are prohibitive for some agents to engage.

In the context of the spread of information, such as news, these graphs could be used to explore strategies to enhance the dissemination of accurate information or to understand the barriers to sharing information. For example, agents with high costs relative to value might represent individuals who are less likely to engage with fact-checking activities or to share information due to high perceived effort or time investment.

The provided equations and the computational process outline a way to model the strategic interactions between agents in the context of an information market, with implications for understanding phenomena like filter bubbles or the spread of fake news. By modeling the value and cost of information, as well as the resulting equilibrium, this approach could offer insights into how to influence agent behavior towards healthier information exchange.

Each pair of graphs corresponds to a different scale of agents (1000, 100, and 10 agents, respectively), indicating that the model has been applied to various-sized populations. This could show how the model scales or how different sizes of agent populations might affect the overall dynamics of the information market.

In order to provide a detailed analysis of the provided network graphs with respect to the fake news spread risk and the maximum compensation issue using the Metzler function.

**Analysis of Minimum Cost Paths for Fake News Spread Risk**

When examining the minimum cost paths within these network graphs, we're considering the cost of spreading information (true or fake) between nodes (news providers or users). The cost is determined based on the truthfulness, importance, and urgency of the information. Each edge in the network graph has an associated cost, which is computed using the Metzler function:

$$C_{ij} = w_1 \times T + w_2 \times I + w_3 \times U$$

where $C_{ij}$ is the cost of spreading information from node $i$ to node $j$, $T$ represents the truthfulness of the information, $I$ the importance, $U$ the urgency, and $w_1, w_2, w_3$ are the respective weights.

**Risk Consideration for Fake News Spread**

High Truthfulness Weight ($w_1$), If the weight for truthfulness is high, it indicates that spreading fake news (higher cost) is penalized within the network. Thus, paths with lower weights may represent channels where verified information is more likely to flow. High Importance and Urgency Weights ($w_2, w_3$), If these weights are significant, it suggests that crucial and timely information bears more "cost" to spread, which might imply a prioritization of such content in the network.

In the context of fake news, a strategy to minimize its spread would involve increasing the cost of spreading fake news between important nodes, effectively "blocking" the most efficient paths for misinformation to proliferate.

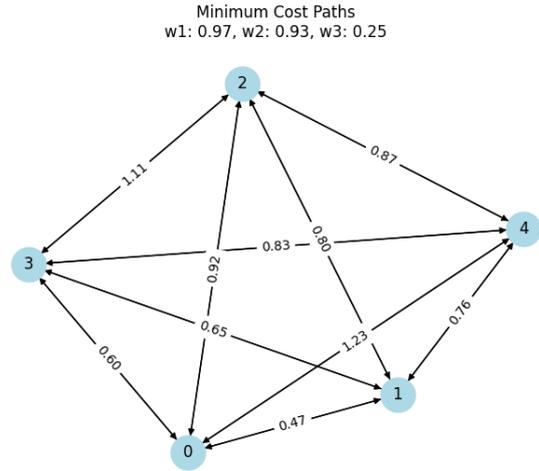

Fig. 17: Minimum Cost Path, Information Value and Cost / Demand-Supply Equilibrium

**Analysis of Maximum Compensation Issue**

The maximum compensation issue refers to the benefits players (or nodes) achieve through cooperative actions, such as sharing accurate information or participating in fact-checking.

**Maximum Compensation Consideration**

Incentivizing Accurate Information Spread, In a network where the maximum compensation issue is addressed, nodes would receive benefits or "compensation" for participating in the dissemination of accurate information. This could be reflected in the network graphs by lower costs for spreading true news. Strategic Adjustments for Compensation, Strategies to enhance information value or reduce costs would be represented by changes in the weights of the Metzler function. For instance, reducing the weight $w_1$ for truthfulness could be seen as a way to lower the "cost" of spreading true news, encouraging its propagation over fake news.

**Considerations Across Different Network Sizes**

When considering the minimum cost paths and maximum compensation issues across networks of different sizes (as seen in the varying number of agents in the graphs), it's essential to note that. Larger Networks, May have more complex interaction patterns and potential pathways for information spread. Thus, strategic interventions may need to be more nuanced to effectively manage fake news risks.

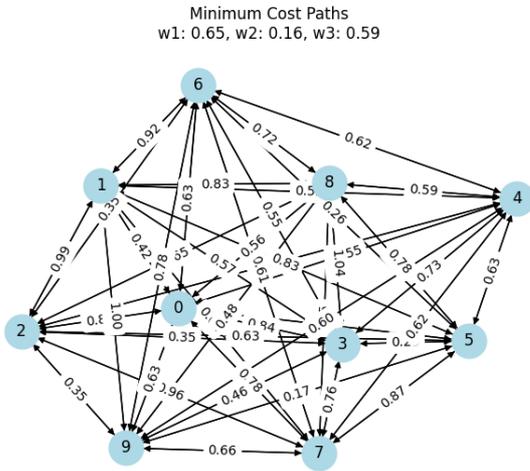

Fig. 18: Minimum Cost Path, Information Value and Cost / Demand-Supply Equilibrium

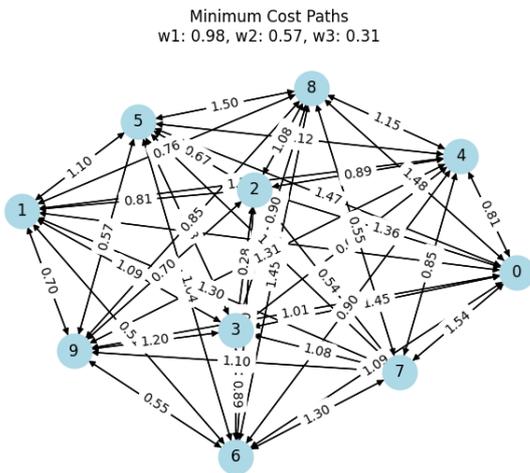

Fig. 19: Minimum Cost Path, Information Value and Cost / Demand-Supply Equilibrium

Results appear to be network graphs with nodes and weighted edges, likely depicting the cost of information transmission between different agents in a network. These graphs part of a study on the spread of information (such as news, including fake news) and how different factors like truthfulness, importance, and urgency affect the cost of spreading information between nodes.

**Analysis Based on Provided Description**

In the context of the provided description, these graphs can be used to demonstrate the application of Metzler functions in a network model to identify minimum cost paths for the dissemination or containment of information. The steps you provided outline how to build a model that can be used to analyze and potentially strategize around the spread of fake news within a network.

**Observations from the Graphs**

Weights on Edges, The numbers on the edges likely represent the cost associated with transmitting information between nodes. These costs are influenced by the weight parameters $w_1, w_2$, and $w_3$, which could correspond to the factors mentioned in the Metzler function definition.

Node Interconnectivity, The denser the network, the more paths information can take. This can affect how quickly and widely information spreads.

**Minimum Cost Paths**

The highlighted paths in the network could represent the most efficient ways to spread information, whether truthful or fake. The model could be identifying these paths based on the lowest aggregate cost from one node to another.

**Strategy Development**

By understanding the costs associated with different paths, strategists can develop interventions to either promote the spread of accurate information or inhibit the spread of misinformation.

**Impact of Parameters**

Changes in the weights $w_1, w_2$, and $w_3$ can significantly alter the cost dynamics in the network, thereby affecting the chosen paths for information transmission.

**Policy Implications**

If this model is used in a real-world context, it can inform policy decisions on how to best allocate resources to fact-checking initiatives or to target specific nodes (e.g., influential social media accounts) for intervention.

Each image seems to represent networks of different sizes, which could be useful for understanding how strategies might scale in different contexts or with different population sizes. The consistency of the approach across different network sizes suggests that the model could be robust across various scenarios.

In summary, these graphs, in conjunction with the outlined steps and equations, represent a complex model that could be used to study the dissemination of information, the cost of spreading fake news, and potential strategies to control or promote certain types of information within a network.

Network diagrams and corresponding weights (w1, w2, w3) represent a model of a communication network where nodes can represent individuals or entities that share information, and edges represent the pathways that information can travel along. The weights on the edges likely represent the cost of information transmission, which could depend on factors such as the reliability of the source (w1), the importance of the information (w2), and the urgency with which the information needs to be disseminated (w3).

**Considering Minimum Cost Paths in the Spread of Fake News**

When considering the spread of fake news, the minimum cost paths would represent the most efficient routes for disseminating misinformation through the network. A strategy to minimize the risk of fake news spread would focus on increasing the costs along these minimum paths. For example, if w1 corresponds to the reliability of the source, strategies could include: Enhancing verification processes to decrease the spread of information from unreliable sources, effectively increasing w1. Improving the literacy of information consumers to recognize unreliable information, which also impacts the effective cost w1.

**Considering the Maximum Compensation Problem**

The maximum compensation problem looks at the opposite end of the spectrum: how to maximize the benefits for cooperative behavior in a network. In the context of the spread of information, cooperative behavior can be seen as the sharing of accurate and reliable information. Strategies here might involve:

Increasing rewards or incentives for sharing verified information, thus reducing the cost of transmission for reliable information and making the truth more competitive in the marketplace of ideas. Building trust and reputation systems that reward nodes for accuracy and punish them for spreading misinformation, altering the cost-benefit analysis for each node.

**Analyzing the Uploaded Graphs:Graph with Weights w1: 0.98, w2: 0.57, w3: 0.31**

This graph likely represents a network with a high emphasis on the reliability of the source (w1 is the highest). The strategy here could involve prioritizing the strengthening of source reliability to increase the cost of spreading fake news.

**Graph with Weights w1: 0.65, w2: 0.16, w3: 0.59**

The emphasis here seems to be on the urgency (w3) and reliability (w1) over the importance of the information (w2). A possible interpretation could be that in scenarios where information needs to be spread quickly (urgent situations), ensuring that the sources are reliable becomes even more critical.

**Graph with Weights w1: 0.97, w2: 0.93, w3: 0.25**

This network places almost equal weight on the reliability and importance of information, with less emphasis on urgency. Such a model suggests that in environments where both the source's reliability and the information's importance are critical, strategies should focus on reinforcing these aspects to increase the information's value and the cost of spreading misinformation.

In all cases, the model assumes that by manipulating the costs associated with these factors, the behavior of the network nodes can be influenced, and the spread of fake news can be either contained or encouraged (for example, in counter-propaganda strategies).

The actual implementation of such strategies would require a thorough understanding of the network dynamics and the ability to influence these weight factors. These models can serve as a basis for designing policies or interventions that aim to improve the overall informational health of a society by promoting the spread of accurate information and reducing the impact of fake news.